\documentclass[a4paper,fleqn]{cas-dc}

\usepackage[authoryear,longnamesfirst]{natbib}
\usepackage[edges]{forest}
\usepackage{xcolor}
\colorlet{hidden-draw}{black}
\def\tsc#1{\csdef{#1}{\textsc{\lowercase{#1}}\xspace}}
\tsc{WGM}
\tsc{QE}
\tsc{EP}
\tsc{PMS}
\tsc{BEC}
\tsc{DE}

\begin{document}
\let\WriteBookmarks\relax
\def\floatpagepagefraction{1}
\def\textpagefraction{.001}
\shorttitle{Leveraging social media news}
\shortauthors{Hao Zhang et~al.}

\title [mode = title]{A Comprehensive Survey on Cross-Domain Recommendation: Taxonomy, Progress, and Prospects}                      



\author[1]{Hao Zhang}[]
\ead{zh2001@mail.ustc.edu.cn}
\author[1]{Mingyue Cheng}[]
\ead{mycheng@ustc.edu.cn}
\author[1]{Qi Liu}[]
\ead{qiliuql@ustc.edu.cn}
\author[1]{Junzhe Jiang}[]
\ead{jzjiang@mail.ustc.edu.cn}
\author[1]{Xianquan Wang}[]
\ead{wxqcn@mail.ustc.edu.cn}
\author[1]{Rujiao Zhang}[]
\ead{zhangrujiao@mail.ustc.edu.cn}
\author[2]{Chenyi Lei}[]
\ead{leichy@mail.ustc.edu.cn}
\author[1]{Enhong Chen}[]
\ead{cheneh@ustc.edu.cn}


\credit{Conceptualization of this study, Methodology, Software}

\affiliation[1]{organization={State Key Laboratory of Cognitive Intelligence, University of Science and Technology of China},   
                city={Hefei,Anhui Province},
                country={China}}
\affiliation[2]{organization={Kuaishou Technology},   
                city={Beijing},
                country={China}}







\begin{abstract}
Recommender systems (RS) have become crucial tools for information
filtering in various real world scenarios. And cross domain recommendation (CDR) has been widely explored in recent years in order to provide better recommendation results in the target domain with the help of other domains. The CDR technology has developed rapidly, yet there is a lack of a comprehensive survey summarizing recent works. Therefore, in this paper, we will summarize the progress and prospects based on the main procedure of CDR, including \textbf{Cross Domain
Relevance}, \textbf{Cross Domain
Interaction}, \textbf{Cross Domain Representation
Enhancement} and \textbf{Model Optimization}. To help researchers better understand and engage in this field, we also organize the applications and resources, and highlight several current important challenges and future directions of CDR. More details of the survey articles are available at \url{https://github.com/USTCAGI/Awesome-Cross-Domain-Recommendation-Papers-and-Resources}.

\end{abstract}




\maketitle

\section{Introduction}
With the rapid development of the Internet, a wide variety of applications and online services have proliferated, leading to an increase in user interaction and data accumulation across different platforms. These datasets span extensive domains, including e-commerce, online video, music services, and social networks, each accumulating substantial information about user preferences and behaviors. To better leverage these multi-domain data, cross-domain recommendation (CDR) systems \citep{fu2024exploring,tang2012cross,singh2008relational,li2009can,li2021cross,zhu2022personalized} have emerged. The core concept of these systems is to integrate and analyze user behaviors and preferences in different domains to provide more accurate and personalized recommendations~\citep{cheng2022towards,jiang2024reformulating,zhang2023towards}. Cross-domain recommendations not only help users discover relevant content across domains, but also enhance user experience and increase the frequency of usage across various applications and services. On the the one hand, cross-domain recommendation breaks the barriers between domains, allowing data and knowledge to flow and be shared across different domains, thereby enhancing the accuracy and user satisfaction of recommendations. However, it can effectively address the issues of data sparsity and the cold start problem that single-domain recommendation systems face. For instance, a new user on a music recommendation platform may lack sufficient behavioral data; however, if active data from a video platform are considered, the system can infer the user's musical preferences and provide more apt music recommendations. Given the promising future of cross-domain recommendation as a significant technology, much research has focused on this area in recent years. To better summarize these efforts, a review and survey are urgently needed.

Generally, traditional recommendation systems aim to infer user preferences through collaborative information~\citep{cheng2021learning,cheng2021softrec}, i.e., historical interaction records, possibly including user profiles and item side information. For simplicity, CDR  can be defined as a recommendation system utilizing information from multiple domains to enhance performance, either of a single aspect or overall. In the following subsections, we will introduce the general procedures and our taxonomy to make the survey more readable.
\subsection{Procedures of CDR}
\begin{figure*}
    \centering
    \includegraphics[width=0.8\textwidth]{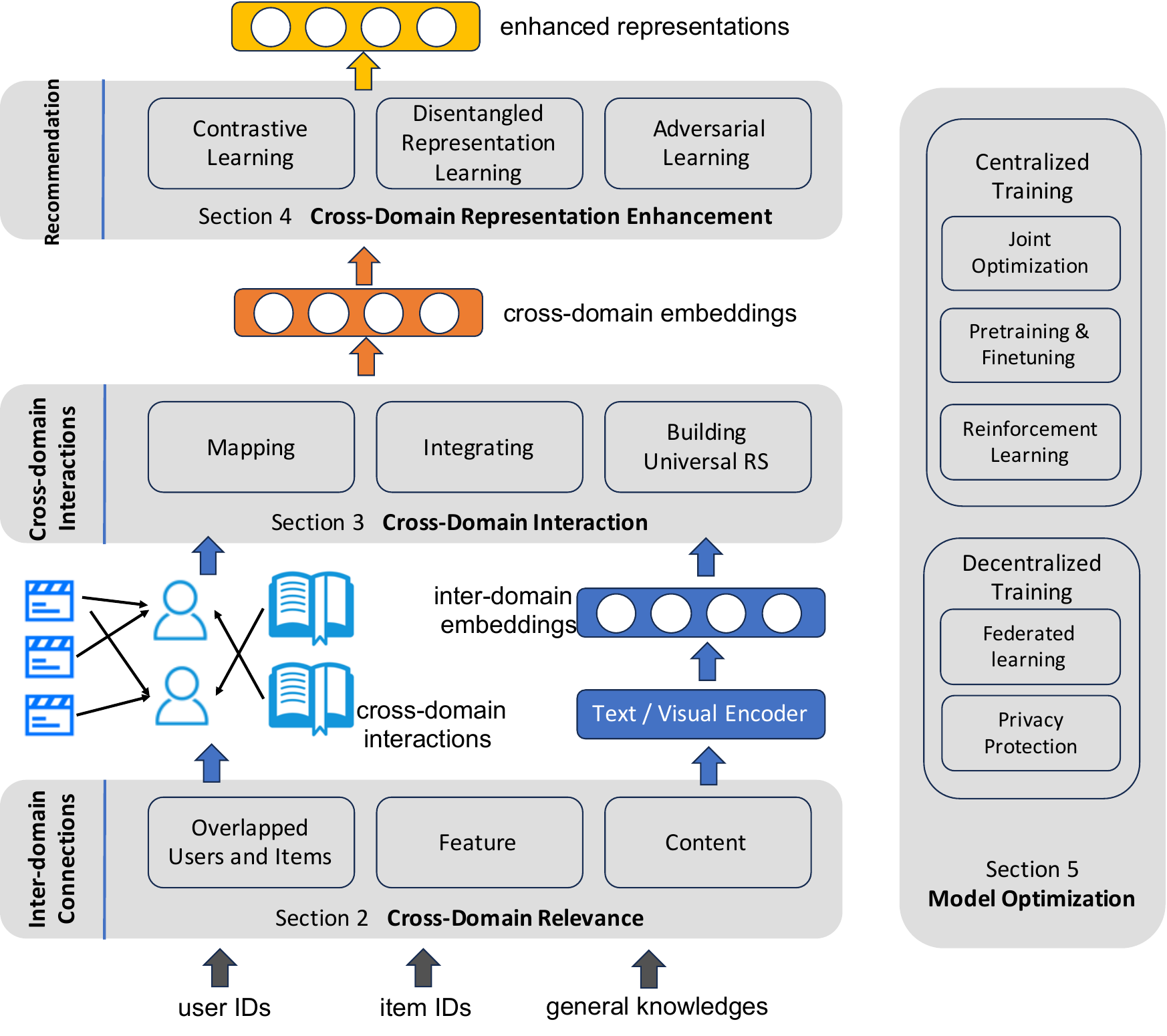}
    \caption{The procedures of cross domain recommender system.}
    \label{fig1}
\end{figure*}

We summarize the general procedures of CDR as Figure \ref{fig1}: Inter-domain Connections, Cross-domain Interaction, and Recommendation. Next, we illustrate this process using the example of a recommendation scenario in e-Commerce.\\
\textbf{Inter-domain Connections.} 
E-commerce recommendation systems typically convert user and item IDs into embeddings, but these embedding layers often cannot be shared directly. Therefore, to leverage information across multiple domains, it is necessary to first establish connections between these domains. This can be achieved by identifying shared elements, such as user-overlapping items across different platforms. Additionally, general knowledge such as product text descriptions and images can also be utilized. By combining these with architectures such as VIT \citep{dosovitskiy2020image} and Bert \citep{kenton2019bert}, tabular features can be extracted for enhanced recommendation capabilities.
\\\textbf{Cross-domain Interactions.} 
The representations of users and items, along with various information,, may exhibit domain-specific characteristics. Therefore, after establishing connections between domains, CDR needs to fuse knowledge from multiple domains to model more accurate user interests and item representations, ultimately achieving better recommendation results~\citep{shen2024pmg}.
\\\textbf{Recommendation.} 
After completing the first two procedures, the models can use knowledge from multiple domains to make recommendations. However, problems such as data sparsity, domain imbalance, and negative transfer still affect model performance. To overcome these challenges, many studies propose to enhance user and item representations by incorporating cross-domain information.



\tikzstyle{my-box}=[
    rectangle,
    rounded corners,
    text opacity=1,
    minimum height=1.5em,
    minimum width=5em,
    inner sep=2pt,
    align=center,
    fill opacity=.5,
    line width=0.8pt
]
\tikzstyle{leaf}=[my-box, minimum height=1.5em,
    text=black, align=left,font=\large,
    inner xsep=2pt,
    inner ysep=4pt,
    line width=0.8pt
]

	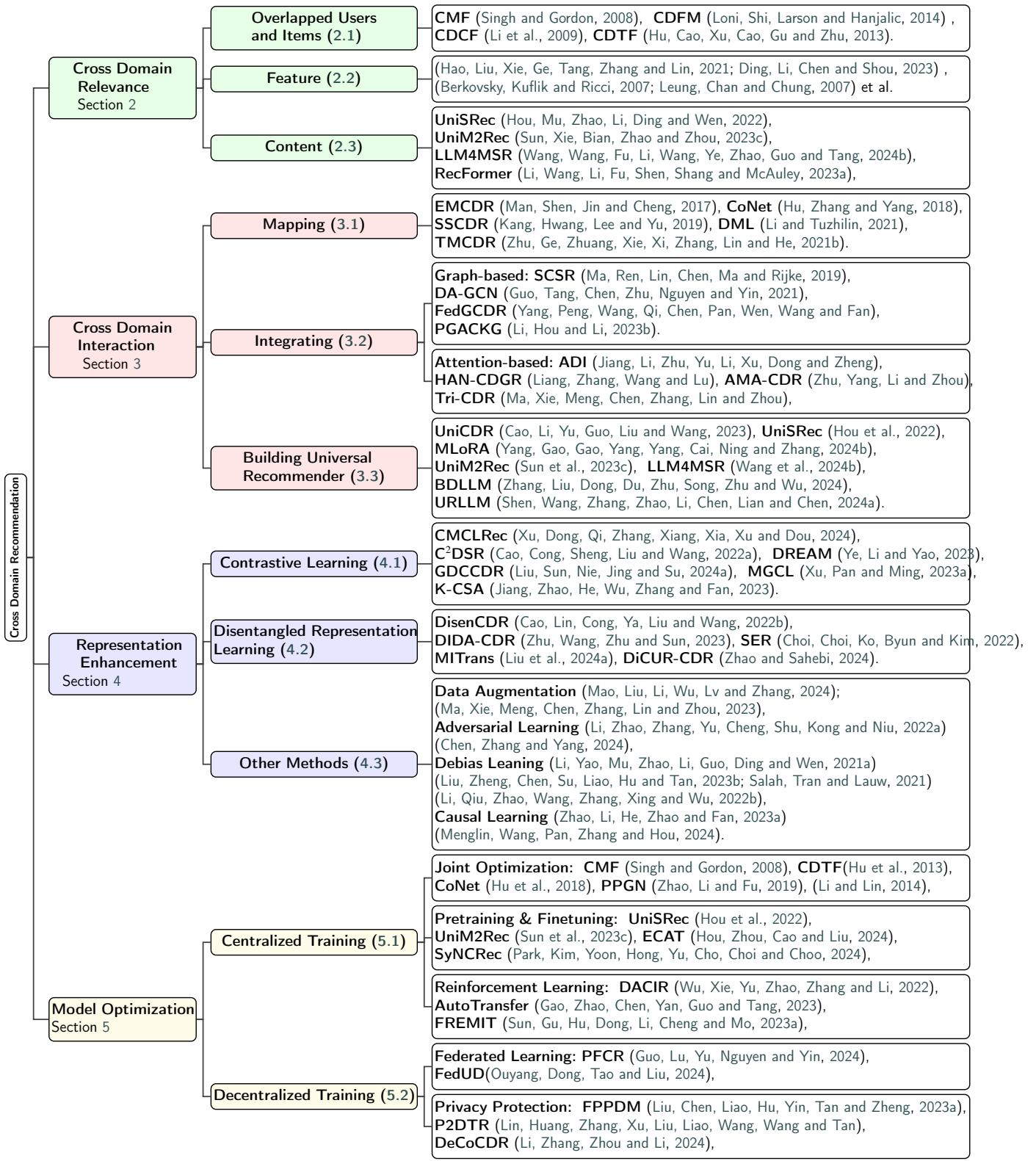
\begin{figure*}[h]
			\centering
            	\resizebox{1\textwidth}{!}{
            \begin{forest}
forked edges,
					for tree={
						grow=east,
						reversed=true,
						anchor=base west,
						parent anchor=east,
						child anchor=west,
						base=left,
						rectangle,
						draw=hidden-draw,
						rounded corners,
						align=left,
						text centered,
						minimum width=4em,
						edge+={darkgray, line width=1pt},
						s sep=3pt,
						inner xsep=2pt,
						inner ysep=3pt,
						line width=0.8pt,
						ver/.style={rotate=90, child anchor=north, parent anchor=south, anchor=center},
					},
					where level=1{text width=14em,font=\large,}{},
					where level=2{text width=15em,font=\large,}{},
					where level=3{text width=17em,font=\large,}{},
					[
					\textbf{Cross Domain Recommendation}, ver
                        [
    					\textbf{Cross Domain} \\ {} \textbf{Relevance} \\ {} Section \ref{rel}, fill=green!10, text width=12em
    					 [ 
    					\textbf{Overlapped Users} \\ \textbf{and Items (\ref{ove})}, fill=green!10, text width=17em
    					 [
             \textbf{CMF}  \citep{singh2008relational}{, } \textbf{CDFM}  \citep{loni2014cross} {,} \\
                        \textbf{CDCF} \citep{li2009can}{,} 
                        \textbf{CDTF} \citep{hu2013personalized}{. }
    					 , leaf, text width=45em
    					]
    					]
                    [
                        \textbf{Feature  (\ref{fea})}, fill=green!10, text width=17em
                        [\citep{hao2021adversarial,ding2023tpuf} {,} \\ \citep{berkovsky2007cross,leung2007applying} et al., leaf, text width=45em]
                    ]
                    [
                        \textbf{Content  (\ref{cont})}, fill=green!10, text width=17em
    					[
					\textbf{UniSRec} \citep{hou2022towards}{,}\\ \textbf{UniM2Rec} \citep{sun2023universal}{, } \\ \textbf{LLM4MSR} \citep{wang2024llm4msr}{, }\\ \textbf{RecFormer} \citep{li2023text}{, }, leaf, text width=45em
    					]
    					]
                    ]
					[
					\textbf{Cross Domain} \\ {} \textbf{Interaction} \\ {  } Section \ref{int}, fill=red!10, text width=12em
					[ 
					\textbf{Mapping (\ref{map})}, fill=red!10, text width=17em
					[
					{\textbf{EMCDR}} \citep{man2017cross}{, }\textbf{CoNet} \citep{hu2018conet}{,
					}\\ \textbf{SSCDR}  \citep{kang2019semi}{,} \textbf{DML}  \citep{li2021dual}{,} \\\textbf{TMCDR}  \citep{zhu2021transfer}{.}, leaf, text width=45em
					]
					]
					[ 
					\textbf{Integrating (\ref{integration})}, fill=red!10, text width=17em
					[      
					\textbf{Graph-based:} \textbf{SCSR} \citep{pinet}{,}\\ \textbf{DA-GCN} \citep{guo_da-gcn_2021}{, } \\\textbf{FedGCDR} \citep{yang_federated_2024}{, }\\ \textbf{PGACKG}  \citep{li2023preference}{.} 
                    , leaf, text width=45em
					]
                    [\textbf{Attention-based:} 
                    \textbf{ADI} \citep{jiang_adaptive_2022}{,} \\
                    \textbf{HAN-CDGR} \citep{liang_hierarchical_2024}{,}
                     \textbf{AMA-CDR} \citep{zhu_active_2024}{,}\\
                      \textbf{Tri-CDR} \citep{ma_triple_2024}{,}
                    , leaf, text width=45em]
					]
					[
					\textbf{Building Universal} \\ 
                    \textbf{Recommender  (\ref{uni})}, fill=red!10, text width=17em
					[\textbf{UniCDR} \citep{cao2023towards}{,}  \textbf{UniSRec} \citep{hou2022towards}{, } \\
					\textbf{MLoRA} \citep{yang2024mlora}{,} \\\textbf{UniM2Rec} \citep{sun2023universal}{, } \textbf{LLM4MSR} \citep{wang2024llm4msr}{, }\\\textbf{BDLLM} \citep{zhang2024bridging}{, } \\\textbf{URLLM} \citep{shen2024exploring}{. },leaf, text width=45em
					]
					]
					]
                    					[
					\;\; \textbf{Representation} \\ \;\;\; \textbf{Enhancement} {  }\\Section \ref{enh}, fill=blue!10, text width=12em
					[
					\textbf{Contrastive Learning (\ref{con})}, fill=blue!10, text width=17em
					[\textbf{CMCLRec} \citep{xu2024cmclrec}{,}\\
					\textbf{C$^2$DSR} \citep{cao2022contrastive}{, } \textbf{DREAM} \citep{ye2023dream}{, } \\ \textbf{GDCCDR }\citep{liu2024graph}{, } \textbf{MGCL} \citep{xu2023multi}{, } \\\textbf{K-CSA }\citep{jiang2023knowledge}{. },leaf, text width=45em
					]
					]
					[
					\textbf{Disentangled Representation} \\ \textbf{Learning (\ref{dis})}, fill=blue!10, text width=17em
					[
					\textbf{DisenCDR} \citep{cao2022disencdr}{, } \\\textbf{DIDA-CDR} \citep{zhu2023domain}{,} \textbf{SER} \citep{choi2022based}{,} \\\textbf{MITrans} \citep{liu2024graph}{,} \textbf{DiCUR-CDR} \citep{zhao2024discerning}{.}, leaf, text width=45em
					]
					]
					[
					\textbf{Other Methods (\ref{oth})}, fill=blue!10, text width=17em
					[
					\textbf{Data Augmentation} \citep{mao2024cross}{;}\\ \citep{ma2023exploring}{,} \\\textbf{Adversarial Learning} \citep{li2022recguru}\\\citep{chen2024improving}{, } \\\textbf{Debias Leaning }\citep{li2021debiasing}\\\citep{liu2023joint,salah2021towards}\\\citep{li2022gromov}{, } \\ \textbf{Causal Learning }\citep{zhao2023sequential}\\\citep{menglin2024c2dr}{.} , leaf, text width=45em
					]
					]
					]
                    					[
					\textbf{Model Optimization} \\  Section \ref{opt}, fill=yellow!10, text width=12em
					[
					\textbf{Centralized Training (\ref{cen})}, fill=yellow!10, text width=17em
                                        [\textbf{Joint Optimization: }  
                   \textbf{CMF} \citep{singh2008relational}{,} \textbf{CDTF}\citep{hu2013personalized}{,} 
                   \\\textbf{CoNet} \citep{hu2018conet}{,} 
                   \textbf{PPGN} \citep{zhao2019cross}{,}
                   \citep{li2014matching}{,}
                    , leaf, text width=45em
                    ]
                    [\textbf{Pretraining \& Finetuning: } 
                   \textbf{UniSRec} \citep{hou2022towards}{,} \\\textbf{UniM2Rec }\citep{sun2023universal}{,}  \textbf{ECAT} \citep{hou2024ecat}{,}\\\textbf{SyNCRec} \citep{park2024pacer}{,}
                   , leaf, text width=45em
                    ]
                    [\textbf{Reinforcement Learning: } 
                    \textbf{DACIR} \citep{wu2022dynamics}{,} \\\textbf{AutoTransfer} \citep{gao2023autotransfer}{,} \\\textbf{FREMIT} \citep{sun2023remit}{,}
                   , leaf, text width=45em
                    ]
                    ]
					[
					\textbf{Decentralized Training (\ref{dct})}, fill=yellow!10, text width=17em
                    [\textbf{Federated Learning:} 
                  \textbf{PFCR} \citep{guo2024prompt}{,} \\\textbf{FedUD}\citep{ouyang2024fedud}{,} , leaf, text width=45em
                    ]
                    [\textbf{Privacy Protection: }
                     \textbf{FPPDM} \citep{liu2023federated}{,} \\\textbf{P2DTR} \citep{linenhancing}{,}  \\\textbf{DeCoCDR} \citep{li2024decocdr}{,}, leaf, text width=45em
                    ]
                    ]
					]
                    ]
\end{forest}}
			\caption{Summary of Cross Domain Recommendation methods.}
			\vspace{-10pt}
			\label{summary}
		\end{figure*}
\subsection{Taxonomy}
\begin{itemize}
    \item Challenge 1: 
    How to establish connections between domains? Due to data isolation on different online platforms, recommendation system models cannot be directly applied across various domains. A common approach is to connect these domains through overlapped users or items. However, many scenarios lack overlapping users or items, which still poses challenges for cross-domain recommendations. Thanks to advancements in computer vision and natural language processing, the use of universal information combined with large language models (LLMs) or multimodal models makes content-based cross-domain recommendations feasible. We summarize the ways of establishing cross-domain connections in Section 2.
    \item Challenge 2: How to effectively utilize cross-domain information and knowledge?
    Modeling cross-domain interactions is crucial for effective recommendations across different domains. The main challenge lies in capturing and leveraging the complex relationships between users and items that span multiple domains. Traditional models often fall short because they are designed for single-domain interactions. To overcome this, new approaches focus on creating unified models that can learn from and adapt to multiple domains simultaneously.
    \item 
    Challenge 3: How to obtain more comprehensive representations for CDR?
Although CDR models have achieved significant success, most methods often ignore that users usually exhibit partial preferences within a specific domain, making the representations practically biased. As a result, directly aggregating representations independently derived from each domain may lead to negative transfer and exacerbate issues related to data imbalance. Compared to traditional CDR, information from other domains should be better utilized for modeling user interests. We introduce this part of the work as Cross-domain representation enhancement in Section 4.
    \item Challenge 4:
    How to optimize the CDR models?
Joint optimization across multiple domains requires careful attention to the performance of each domain, which can increase model complexity and be susceptible to domain imbalance. Additionally, privacy protection is crucial. As data privacy awareness grows, techniques like federated learning are being adopted to safeguard data, placing higher demands on CDR models design.
\end{itemize}
We organize the paper according to the main challenges of CDR as mentioned above. 
We also noted two existing surveys. One \citep{zhu2021cross} focus on the training targets of cross-domain recommendation, and the other one \citep{zang2022survey} organize the research according to the user-item overlap scenarios. Different from previous surveys, we summarize past work from a technical perspective based on cross-domain recommendation procedures, and the corresponding representative works is shown in Figure \ref{summary}. Besides, this article also includes the latest advancements such as LLM-based approaches and cross-domain representation enhancement methods, which are not covered in earlier surveys.

\section{Cross-Domain Relevance} \label{rel}
\subsection{Overlapped Users and Items}\label{ove}

User-item interaction data is the core of recommendation systems, typically including users' historical behavior data such as clicks, purchases, ratings, or viewing records. In cross-domain recommendation systems, these interactions are not limited to a single domain. For example, a user's viewing history in a movie recommendation system might help predict their preferences in a music recommendation context. By analyzing user interaction patterns across various domains, the system can identify overlapped users' broad interests and preferences, thereby enabling more accurate personalized recommendations using data from one domain in another \citep{singh2008relational,loni2014cross,li2009can,hu2013personalized}.

\subsection{Feature}\label{fea}

Feature information plays a crucial role in building recommendation systems. These features may include demographic information about users (such as age, gender, geographical location), device usage, session duration, as well as attributes of items like category, price, brand, and content description. In cross-domain scenarios, features from different domains can be integrated to discover correlations and preference patterns in user behavior across domains \citep{hao2021adversarial,ding2023tpuf}. For example, a preference for a specific genre (like science fiction novels) in the book domain might suggest a preference for science fiction movies in the film domain.

\subsection{Content}\label{cont}
In many cross-domain recommendation scenarios, there may be almost no overlapping users or items between different domains. Therefore, in these scenarios, we may need to establish more general connections through content information.
Content information refers to information directly related to items, such as text descriptions, images, videos, or audio content. In cross-domain recommendation systems, content information can be used to extract cross-domain features, which help link similar items or user behaviors across different domains. For instance, by analyzing the descriptive texts of movies and books, similarities in themes or sentiments can be identified, which can then be used to recommend books with similar themes to users who like corresponding movies \citep{hou2022towards,li2023text}.

\section{Cross-Domain Interaction}  \label{int}
In cross-domain recommendation, it is crucial to measure the interaction signals between users and between users and items.
We classify Cross-Domain Interaction methods into three categories: mapping, integration, and building universal recommender system.

\subsection{Mapping} \label{map}
The Mapping method~\citep{gao2019expert}, as its name suggests, aims to learn a mapping function from the source domain to the target domain. The fundamental assumption of this approach is that there \textbf{exists a mapping relationship} between the feature vectors of users/items in the source and target domains. The main concern with previous methods is that simultaneously learning both domain-specific and shared factors may exacerbate data sparsity.

As a pioneer of the Mapping-based approach,~\cite{finn2017model} introduced a general and model-agnostic meta-learning algorithm aimed at enabling artificial agents to learn quickly and adaptively from limited data, akin to human intelligence. This method focuses on optimizing the initial parameters of deep neural network models so they can rapidly learn new tasks after only a few gradient updates with minimal new data. Unlike previous meta-learning approaches that modify model architectures or learning rules, this algorithm remains architecture-agnostic, allowing it to be applied to various neural network structures, including fully connected, convolutional, and recurrent networks. The key innovation lies in training the model to maximize its performance on new tasks by ensuring that small adjustments to parameters lead to significant improvements in task-specific loss, thereby enhancing sensitivity to changes. This results in a robust internal representation suitable for diverse tasks, facilitating rapid fine-tuning and effective learning across various domains such as few-shot regression, image classification, and reinforcement learning. The algorithm demonstrates superior performance compared to state-of-the-art one-shot learning methods, achieving comparable or better results with fewer parameters and significantly enhancing reinforcement learning efficiency in variable task scenarios. Overall, this work provides a versatile framework for fast and effective meta-learning applicable across a wide range of machine learning problems.

For example, EMCDR~\citep{man2017cross} addresses the data sparsity problem in recommender systems with cross-domain recommendation. They investigate two key questions: whether to use linear or nonlinear mapping functions between domains, and which subset of data should be used for learning these mappings. They specifically aim to improve upon existing approaches which either work asymmetrically (using auxiliary domain data to help the target domain) or symmetrically (treating both domains equally and learning domain-specific and domain-sharing factors). 
CoNet~\citep{hu2018conet} addresses these challenges by enabling knowledge transfer between multiple domains through dual mappings, allowing hidden layers of neural networks from different domains to interact and share information effectively. By incorporating dual shortcut connections and joint loss functions, CoNet enhances the ability to learn complex user-item interactions while improving recommendation performance.

Based on this, \cite{cui2020herograph} introduced a Heterogeneous GRAPH framework for Multi-Target Collaborative Filtering (HeroGRAPH), aimed at addressing the data sparsity issue in recommender systems through Multi-Target Cross-Domain Recommendation. Unlike previous methods that primarily focus on Single-Target (STCDR) and Dual-Target Collaborative Filtering (DTCDR), HeroGRAPH leverages a shared graph structure that integrates user and item data across multiple domains, allowing for simultaneous modeling of within-domain and cross-domain behaviors.

Moreover, ~\cite{kang2019semi} introduced several innovative contributions to enhance recommendation systems for cross-domain recommenders. It employs a semi-supervised learning approach that utilizes both labeled data from overlapping users and unlabeled data from user-item interactions, effectively addressing the cold-start problem. By learning latent vectors within metric spaces, SSCDR captures user-user and user-item similarities more accurately than traditional models that rely on inner products, which can distort relationships due to violations of triangle inequality. Additionally, it trains a cross-domain mapping function that optimizes for both the preferences of overlapping users and the interactions of cold-start users, resulting in improved accuracy in inferring user preferences. Notably, SSCDR demonstrates strong performance even in scenarios with minimal overlapping users, reflecting realistic user distributions, and employs a novel technique to accurately infer the latent vectors of cold-start users by considering their neighborhoods.

Then, ~\cite{li2021dual} integrate metric learning. The proposed DML identifies not only overlapping users but also those with similar preferences, significantly reducing the reliance on a large number of overlapping users, which can be challenging to gather. The method learns a joint latent metric space that captures user preferences and similarities among users and items, enabling more accurate identification of relationships in user preferences. Additionally, DML employs an iterative process to update the model, simultaneously improving recommendation performance across both domains. Empirical validation demonstrates that DML consistently outperforms state-of-the-art approaches, showcasing its effectiveness in cross-domain recommendations even with minimal user overlap. Overall, DML presents a robust solution by effectively leveraging dual learning and metric learning to optimize performance in diverse recommendation scenarios.

Furthermore, ~\cite{zhu2021transfer} present a novel Transfer-Meta CDR (TMCDR) framework to tackle the cold-start problem in recommender systems, which often struggle with new users and items. By leveraging cross-domain recommendation (CDR), TMCDR addresses limitations of existing methods, particularly the Embedding and Mapping approach (EMCDR), which relies heavily on overlapping users for learning mapping functions, leading to biases and poor generalization. TMCDR consists of two stages: a Transfer stage that utilizes pre-trained models from each domain to enhance the embedding process, and a Meta stage that develops a task-oriented meta network to facilitate the transformation of overlapping user embeddings to the target domain's feature space. This meta network is optimized using Model-Agnostic Meta-Learning (MAML) and is designed to improve performance on rating or ranking tasks for cold-start users. The framework is versatile, applicable to various base models like Matrix Factorization (MF), Bayesian Personalized Ranking (BPR), and Conditional Markov Linkage (CML), and is particularly effective in online cold-start settings, enabling immediate application for new users.

\subsection{Integration}  \label{integration}
Integration, in essence, refers to the process of combining and learning interaction information from multiple domains and different granularities to enable a more comprehensive and nuanced modeling of user preferences. Specifically, "integration" involves the fusion of multi-dimensional data from various domains, enhancing the accuracy and generalization capability of recommendation systems through cross-domain, cross-sequence, and even intra-sequence interactions.

\subsubsection{Graph Neural Network Based Methods}  \label{gnn}
In practice, there are many graph-based approaches. As pioneers in this field, some researchers~\citep{cui_herograph_2020} proposed a method that leverages graphs to model cross-domain recommendation problems, especially when addressing the modeling of the "novelty-seeking trait." Their approach utilized graph structures to represent relationships between different domains, thus facilitating knowledge transfer. However, their work overlooked important sequential information and relied on explicit user ratings, which are often unavailable across both domains. Furthermore, these cross-platform methods were not designed to handle intertwined user preferences under shared accounts. One of the prior works in shared-account cross-domain recommendation (SCSR) is the $\pi$-net method~\citep{pinet}, which frames SCSR as a parallel sequence recommendation problem and resolves it using an information-sharing network. Another relevant work is the PSJNet~\citep{psjnet}, which improves upon $\pi$-net by employing a split-connection strategy. However, these RNN-based methods primarily focus on capturing sequential dependencies and are limited in their ability to capture complex relationships between associated entities (i.e., users and items) across domains. This limitation restricts the expressiveness of the learned user and item representations and ignores the explicit structural information that connects the two domains, such as item-user-item paths.

To address these challenges, a novel graph-based solution called the Domain-Aware Graph Convolutional Network (DA-GCN)~\citep{guo_da-gcn_2021} is proposed for shared-account cross-domain recommendation (SCSR). Specifically, to model the complex interaction relationships, they first construct a cross-domain sequence (CDS) graph that links different domains, where users and items from each domain are represented as nodes, and their relationships are modeled as edges. To adapt to the dispersed user preferences on shared accounts, the model assumes the existence of some potential users per account and utilizes an information propagation strategy within the domain-aware graph convolutional network to aggregate information from directly connected neighbors. By propagating information from these neighboring nodes, the model learns node embeddings for both users and items. While interactions with items are recorded at the account level, the attraction of items from different domains to different users within the same account varies. Therefore, two specialized attention mechanisms are further designed to selectively choose relevant information during the information propagation process. Ultimately, by considering structural information, the proposed method is able to model multifaceted interaction relationships and effectively propagate refined domain knowledge.

Apart from them, some researchers have also explored the integration of knowledge graphs to collaboratively handle cross-domain interaction information. For instance, a novel approach was proposed that utilizes a preference-aware graph attention model~\citep{li2023preference}, which incorporates a collaborative knowledge graph for cross-domain recommendations. This model is designed to capture both entity semantic information and users' long-distance preferences across different domains. Specifically, it leverages a trainable and personalized graph representation scheme to transform entities or items into preference-aware embeddings. Addressing the limitations of existing solutions, a preference-aware graph attention network is introduced to aggregate preference features of similar entities both within and across domains. Finally, the fused entity features, enriched with contextual preference information, are used in a cross-domain Bayesian personalized ranking method to generate predictive results for cross-domain recommendation tasks.

Recent work in graph-based models has introduced new innovations to address privacy concerns. To tackle the challenges of privacy protection and negative transfer in cross-domain recommendation, a novel approach called Federated Graph Learning for Cross-Domain Recommendation (FedGCDR)~\citep{yang_federated_2024} is proposed. 
The first module of it, aims to protect inter-domain privacy and reduce potential negative transfer before the knowledge transfer process. This module leverages differential privacy~\citep{dwork2014algorithmic}, ensuring theoretical privacy guarantees, and aligns feature spaces to facilitate effective knowledge transfer.
The second module is the Positive Knowledge Activation Module, which is designed to further mitigate negative transfer. Specifically, it enhances the local graph of the target domain by incorporating virtual social links, which facilitates the generation of domain-specific attention mechanisms.

\subsubsection{Attention Mechanism Based Methods}  \label{attn}
In cross-domain scenarios, it is a natural idea to assign higher weights to important domain information in order to obtain more practical insights. Existing Cross-Domain Recommender Systems (CDRSs) are typically developed for individual users and cannot be directly applied to group recommendations. To address the data sparsity issue in Group Recommendation Systems (GRSs), ~\cite{liang_hierarchical_2024} have investigated the cross-domain group recommendation problem and proposed a hierarchical attention network-based method called HAN-CDGR. HAN-CDGR leverages data from a source domain to enhance recommendation generation for both individual users and groups in the target domain, where data sparsity prevents the generation of accurate recommendations. The method constructs a hierarchical attention network to model both individual and group preferences, taking into account group members' interactions, dynamic weights, and the complex relationships between individuals and groups. Additionally, adversarial learning is employed to effectively transfer knowledge from the source domain to the target domain. 

Further research has uncovered additional issues with existing methods. While existing efforts in multi-domain recommendation systems~\citep{sheng2021one} primarily focus on the ranking step, the retrieval step has been less explored~\citep{shen2021sar}. To address the challenges in multi-domain recommendation during the retrieval step, researchers proposed the Adaptive Domain Interest (ADI) network~\citep{jiang_adaptive_2022}, which simultaneously learns users' preferences across multiple scenarios. The ADI model is designed to capture both commonalities and diversities across different domains by employing common networks and domain-specific networks. To tackle feature-level domain adaptation, two domain adaptation techniques are introduced: domain-specific batch normalization and a domain interest adaptation layer. Moreover, to capture label-level connections between domains, a self-training method is incorporated, enhancing the model's ability to transfer knowledge across domains.

Furthermore, ~\cite{zhu_active_2024} have also attempted to incorporate mask mechanisms into attention-based cross-domain methods. In order to enhance the exchangeability of knowledge among multiple domains, they propose constructing interaction graphs for different domains and integrating these graphs using a suitable mask strategy. This approach significantly improves knowledge exchange across domains. The idea behind this is that embedding propagation and combination for common users can help uncover underlying relationships between different graphs (or domains). 
However, a critical research problem in the literature severely limits the effectiveness of existing Cross-Domain Recommender (CDR) approaches. The problem is negative transfer caused by undifferentiated knowledge transfer. As the number of auxiliary domains increases, without a well-designed mechanism to mitigate the negative influence of auxiliary domains, negative transfer becomes almost inevitable. Some methods~\citep{liu2019loss, ge2014handling,wang2019characterizing} have attempted to tackle this issue, they often overlook important transfer-related prior knowledge, such as data density and sample uncertainty, which are crucial for effective Cross-Domain Recommendations.

Beyond this, ~\cite{ma_triple_2024} have explored the combination of triple sentences and attention mechanisms for cross-domain recommendation. This framework is model-agnostic and aims to jointly model the source, target, and mixed sequences in Cross-Domain Sequential Recommendation (CDSR). Specifically, three sequence encoders are built to process the source, mixed, and target domains separately, capturing intra-domain behavior interactions to generate three hidden sequence representations.
To mitigate the issue of irrelevant negative transfer, a Triple Cross-Domain Attention (TCA) method is designed to operate on these three sequence representations. The TCA captures informative knowledge related to users' target-domain preferences and global interests. The attention-enhanced sequence representations are then combined and passed through a Multi-Layer Perceptron (MLP) to generate the final user representation.
In addition, a Triple Contrastive Learning (TCL) strategy is introduced to comprehensively model the correlations among the three sequences. TCL applies three contrastive learning losses to capture coarse-grained similarities between any two sequence representations of the same user, as compared to other users. More importantly, TCL also incorporates a margin-based triple loss to model fine-grained distinctions among the three sequences, ensuring that the information diversity across domains is preserved.

With the continuous improvement of attention mechanisms, how to more efficiently harness their powerful capabilities to empower cross-domain recommendation remains an important area for further research.


\subsection{Building Universal Recommender System}  \label{uni}
Existing cross-domain recommendation approaches introduce specific-domain modules for each domain, which partially address these issues but often significantly increase model parameters and lead to insufficient training. To this end, \cite{yang2024mlora} introduce multiple LoRA (MLoRA) networks where a specialized LoRA module is conducted for each domain. Besides, \cite{cao2023towards} highlight a universal perspective for CDR, and propose a a unified model UniCDR which can transfer the domain-shared information in diverse CDR scenarios by a single model.

Despite effectivess, these methods mainly transfer based on collaborative signals or user-item interactions, may still face substantial challenges in scenarios with no overlapping users or items. With the advancement in computing power, training a universal recommendation model using more general content information has emerged as another effective solution.
To overcome the difficulty of transferring to new recommendation scenarios, UniSRec \cite{hou2022towards} utilizes the associated description text of items to learn transferable representations across different recommendation domains. Based on which, mixture-of-experts (MoEs) enhanced adapter and contrastive pre-training tasks are designed to learn both item and sequence representations respectively. Similarly, \cite{li2023text} formulate items as key-value attribute pairs for the ID free sequential recommendation. Then, the model can transfer item knowledge into different or cold-start recommendation domains and recommend based on language representations. Furthermore, \cite{sun2023universal} bring in the consideration of multi-modal item contents and user preferences from all interacted domain, proposing the first pre-trained multi-domian-recommendation model which introduce the textual, visual and cross-modal MoEs to enhance the robustness of item content representations.

Large language model (LLM) is known for its remarkable capability of language understanding and reasoning in recent years. \cite{tang2023one} explore the ability of language models in multi-domain
behaviors modeling. The experimental results demonstrate
the effectiveness of LLM and the multi-domain information and some interesting findings about model size and zero-shot scenarios are also provided. Knowledge is pre-encoded in LLMs, facilitating the bridging of different domains and enabling the seamless delivery of personalized suggestions across domains to users.
To this end, \cite{petruzzelli2024instructing} designed a pipeline to instruct and prompt LLMs for
explainable cross-domain recommendations. Similarly,  LLM4MSR \citep{wang2024llm4msr} designed scenario- and
user-level prompt without fine-tuning the LLM to
explicitly improve the scenario-aware and personalized recommendation capability. In addition, to bridge the information gap between domain-specific model
and LLM, BDLM \citep{zhang2024bridging} design an information sharing module to transfer information between
domain-specific models and LLMs for RS task, while URLLM \citep{shen2024exploring} develops a user retrieval bounded interaction paradigm which can integrate structural-semantic and
collaborative information into LLM in a seamless manner.

In summary, the unified cross-domain recommender system can integrate data resources from multiple domains through a single model framework, significantly reducing model complexity. By utilizing general content information, such as textual and visual data, these models can further uncover potential connections between different domains, even in the absence of overlapping users and items. These approaches enhance the ability to provide more accurate and personalized recommendations to users while maintaining broad applicability across various contexts.

\section{Cross-Domain Representation Enhancement}  \label{enh}
In the real world, users usually exhibit partial preference involved in a specific domain, making the representations practically biased. Directly aggregating representations independently derived  from each domain may lead to negative transfer and exacerbate issues related to data imbalance. This imbalance might cause certain domain information to be overlooked or overemphasized during integration. To address these challenges, researchers have recently proposed methods mainly based on \textbf{Disentangled Representation Learning (DRL) }and \textbf{Contrastive Learning (CL)}. These methods effectively enhance the model's adaptability and robustness for CDR by aligning representations or separating unique and common characteristics.
\subsection{Contrastive Learning}  \label{con}
Contrastive learning methods have gained significant attention recently to address issues like data imbalance and user activity inconsistency for cross-domain recommendation (CDR). 

Some studies focus on tackling cold-start and data imbalance challenges. The CMCLRec ~\cite{xu2024cmclrec} framework generates simulated behavior sequences and employs self-supervised training to enhance recommendation performance for cold-start users, effectively mitigating the cold-start problem. Concurrently, some research emphasizes enhancing CDR robustness through contrastive learning, particularly addressing user activity inconsistency caused by data imbalance \cite{yang2024not}.

In terms of user and item representation, the C$^2$DSR model captures intra- and inter-sequence item relationships to obtain both single-domain and cross-domain user representations, introducing a contrastive infomax objective to enhance representation relevance \cite{cao2022contrastive}. 
Considering the inter-user sequence relationship, DREAM \cite{ye2023dream} further leverages sequence augmentation to maximize the relevance among inter-sequence with the same preferences and minimize the relevance among inter-sequence with different preferences. 
The GDCCDR \cite{liu2024graph} model introduces disentangled graph updates, leveraging contrastive learning constraints to enhance disentanglement by focusing on domain-invariant and domain-specific features.

Additionally, researchers have developed a unified cross-domain heterogeneous graph to capture cross-domain similarity between entities via a novel message-passing mechanism, employing contrastive learning and gradient alignment to ensure user interest consistency \cite{zhao2023cross}. The MGCL framework \cite{xu2023multi}  addresses the problem from intra-domain item representation and inter-domain user preference perspectives, using contrastive mechanisms to capture dynamic sequential information, static collaborative information, and transition patterns across different domains.

Moreover, K-CSA  \cite{jiang2023knowledge} introduces class semantic descriptions (CSDs) modeling for the first time for cross-domain zero-shot recommendation. This approach aligns items through a knowledge graph and utilizes multi-view K-Means and self-attention mechanisms to extract user CSDs, enhancing cross-domain alignment through cross-semantic contrastive learning. And one study pioneers the application of contrastive learning in CDR matching by designing cross-domain contrastive learning tasks through subgraph data augmentation and diversified preference networks \cite{xie2022contrastive}, effectively enhancing the diversity and efficiency of knowledge transfer paths.

These studies demonstrate the potential of contrastive learning to enhance the performance of cross-domain recommendation systems, providing important directions and insights for future research.
\subsection{Disentangled Representation Learning}  \label{dis}
Existing CDR methods often fail to disentangle domain-shared information from domain-specific information, directly aggregating these two aspect information may lead to suboptimal performance. Luckily, in recent year, a series works have been proposed to adopt solve CDR with disentangled representation learning. These approaches often disentangles domain-shared and domain-specific information, fusing only the domain-shared information to enhance transfer effectiveness while avoiding the negative transfer of irrelevant information.

DisenCDR \citep{cao2022disencdr} pioneers the use of mutual information-based regularizers to learn disentangled representations, distinguishing between domain-specific and domain-shared preferences, thus improving the model's recommendation performance.

To further emphasize augmenting the sparser target domain, DIDA-CDR \citep{zhu2023domain} proposes a disentanglement-based framework combined with interpolative data augmentation to generate more diverse user representations, thereby enhancing recommendation accuracy. As for the scenario even when users or items are not shared between domains, SER \citep{choi2022based} adaptively disentangles features depending on the characteristics of the source and target domains, make it applicable to heterogeneous scenarios, in case the source and target domain have less similar
characteristics. Similarly, MITrans \citep{liu2024graph} introduce a novel upper bound and a
novel lower bound for mutual information (MI) to explicitly measure the MI
between user preferences in different domains to learn both domain-specific and domain-shared preferences without requiring overlapped users or items.

Besides, methods such as DiCUR-CDR \citep{zhao2024discerning} introduce DisCCA, a novel canonical correlation analysis-based technique, to facilitate the exchange of preferences between domains while acknowledging the unique variations in each domain. This approach ensures both domain-shared and domain-specific preferences are properly considered during recommendation generation. And $C^2DR$ \citep{menglin2024c2dr} incorporates a causal perspective by explicitly identifying domain-specific and domain-shared information as independent variables within a causal graph, enhancing the model's robustness and transfer learning capabilities through disentanglement regularization terms.

Other notable contributions focus on enhancing recommendation through graph-based approaches. For example, GDCCDR \citep{liu2024graph} introduces a disentangled graph update mechanism, employing contrastive learning to enhance the disentanglement of domain-invariant and domain-specific features. The use of random walk-based domain alignment strategies, as seen in MDR \citep{ning2023multi} , also contributes to enhancing cross-domain knowledge sharing by identifying similar users and items across domains.

In summary, these approaches highlight a comprehensive range of techniques for disentangling domain-specific and domain-shared information, addressing challenges such as sparse data and multi-domain alignment. By refining the ability to transfer relevant knowledge across domains, these models significantly advance the accuracy and applicability of cross-domain recommendation systems.


\subsection{Other Enhancement Methods}  \label{oth}
In addition to the aforementioned methods, numerous other efforts have been employed to enhance representation learning in cross-domain recommendation (CDR). 

Some studies approach from a data perspective. \cite{mao2024cross} present CrossAug, a novel data augmentation approach that applies a cross-domain graph propagation module to learn representations of users and items and leverage interactions in two domains more efficiently. \cite{ma2023exploring} explores the phenomenon of false hard negative samples (FHNS) and proposes both general and cross-domain real HNS selectors in CDR. These methods mitigate negative transfer in CDR and identify its corresponding FHNS via a dynamic user-based FHNS filter to maintain its effectiveness.

\textbf{Causal learning} has emerged as a promising approach in cross-domain recommendation by explicitly modeling and leveraging causal relationships to mitigate spurious correlations and improve generalization. ARISEN \citep{zhao2023sequential} integrates causal reasoning into its framework to effectively eliminate the influence of interfering factors, thereby enhancing recommendation performance. Furthermore, it employs the collaborative optimization of instrumental variables, avoiding the limitations of manual definitions and improving the framework’s generalization ability.  
Building on the causal perspective, C\textsuperscript{2}DR \cite{menglin2024c2dr} introduces causal graphs and multiple encoders to model causal relationships in the cross-domain recommendation process. Unlike direct merging of domain-shared and domain-specific information, this approach separates them, improving model performance. Additionally, by using sample weighting techniques and classifiers, C\textsuperscript{2}DR further refines this separation, leading to enhanced accuracy and robustness in recommendations.

Besides, some studies utilize \textbf{adversarial learning} to mitigate data sparsity and enhance the robustness of CDR. RecGURU \citep{li2022recguru} introduces a domain discriminator and a self-attentive auto-encoder to derive latent user representations. Then, an adversarial learning method is then employed to train the two modules, aiming to unify user embeddings generated from different domains into a single global generalized user representation for each user. Similarly, \cite{chen2024improving} proposes Cross-domain Distributional Adversarial Training (CDAT), which utilizes a richer auxiliary domain to improve the adversarial robustness of a sparse target domain. The network processes a domain-invariant preference distribution as input and captures a high-quality Cross-Domain Adversarial Example (CDAE) distribution to support CDAT.

Considering the challenge of domain bias, \cite{li2021debiasing} is the first to apply \textbf{debiasing learning} to CDR, which develops a causal-based approach to mitigate this problem. They correct data selection bias in cross-domain scenarios using a generalized propensity score and further model user preference shifts across different domains by employing propensity scores to address preference bias. 
Moreover, many approaches also employ domain alignment \citep{liu2023joint,li2022gromov,salah2021towards} to reduce domain bias. 
UDMCF \citep{liu2024user} introduces an innovative approach that combines imbalanced distribution optimal transport with a typical subgroup discovery algorithm to map user distributions. This method effectively aligns user representations by addressing distribution disparities, facilitating more accurate recommendations in cross-domain scenarios. Building on the foundation of optimal transport, MOTKD \cite{yang2023multimodal} further enhances cross-domain recommendation by employing a proxy distribution and Sinkhorn-based matching for cross-domain knowledge transfer. Additionally, hierarchical knowledge distillation leverages richer source domain data to improve target domain predictions, providing a complementary mechanism to optimal transport for refining recommendation performance.
The core idea of this strategy is to diminish domain discrepancies and transfer useful knowledge across domains to enhance model performance.

\section{Model Optimization} \label{opt}
\subsection{Centralized Training} \label{cen}
In the realm of cross-domain recommendation, model optimization is crucial for enhancing performance. Centralized training, as a significant approach in this process, has attracted much attention. It aims to optimize models in a unified and coordinated manner, enabling them to better utilize data from multiple domains and improve generalization ability. This section delves into the three main aspects of centralized training, namely joint optimization, pretraining \& finetuning and reinforcement learning, exploring their mechanisms, advantages, and potential challenges.
\subsubsection{Joint Optimization}
The joint training method achieves cross - domain recommendation by training the model simultaneously in the source domain and the target domain. The key of this method lies in designing a shared model structure that can be applied to multiple domains, as well as an appropriate objective function, enabling the model to learn the knowledge of multiple domains collaboratively during the training process. For example, CMF \citep{singh2008relational} decomposes multiple interaction matrices simultaneously and shares parameters among various domains; CDTF \citep{hu2013personalized} proposes a ternary decomposition method to learn the ternary relationships among users, items, and domains; \cite{li2014matching} use the nearest - neighbor search algorithm to match users and items in the latent space; CoNet \citep{hu2018conet} employs an MLP as the basic encoder for each domain and designs a cross - connection network to transfer information during co - training; PPGN \citep{zhao2019cross} uses multiple stacked Graph Convolutional Networks (GCN) to aggregate representation information from various domains for learning user and item representations.

The joint training method can utilize data from multiple domains for training simultaneously, which improves the generalization ability of the model. However, it requires careful design of the model structure and the objective function to balance the task weights among different domains, avoid the excessive influence of one domain on model training, and is vulnerable to the negative transfer phenomenon (that is, some information from the source domain reduces the performance of the model in the target domain instead). 

\subsubsection{Pretraining \& Finetuning}
In cross-domain recommendation, pre-training and fine-tuning offer the advantage of transferring generalizable knowledge from one domain to another in a scalable, efficient manner. During the pre-training phase, models are trained on a broader dataset, such as single-domain data or multi-modal and multi-domain interactions, to learn transferable representations that capture general patterns of user behavior, item semantics, and domain relationships. This phase allows the model to build a robust foundation, even in scenarios with noisy or incomplete data. The fine-tuning phase tailors the pre-trained model to the specific characteristics of the target domain. By focusing on cross-domain data, fine-tuning adapts the learned representations to improve relevance and accuracy in the target recommendation tasks.

For example, UniSRec \citep{hou2022towards} addresses the challenge of transferring learned representations across different recommendation scenarios by utilizing the associated description text of items to learn transferable representations. It employs a lightweight item encoding architecture based on parametric whitening and mixture-of-experts enhanced adaptor.
When considering multimodal scenarios, UniM2Rec \citep{sun2023universal} presents an innovative approach for multi-modal multi-domain recommendation that considers both item contents and user preferences from all interacted domains during pre-training, which is crucial for real-world scenarios where data may be noisy or incomplete.

Additionally, the working mechanism of some two-stage models is also similar to the pre-training and fine-tuning paradigm, SyNCRec \citep{park2024pacer} can be considered that the proposed adaptive weight adjustment and mutual information maximization methods are similar to the process of pre-training and fine-tuning. During the pre-training phase, the model trains on single-domain data to build a model for single-domain features; and during the fine-tuning phase, the model fine-tunes on cross-domain data to improve cross-domain recommendation performance.
Considering efficiency issues and more refined knowledge transfer,  ECAT \citep{hou2024ecat} consists of two core components: first, for sample transfer, the authors proposed a two-stage method that realizes a coarse-to-fine process. Specifically, they performed an initial selection through a graph-guided method, followed by a finer selection using domain adaptation method. Second, they proposed an adaptive knowledge distillation method for continuously transferring representations from a model trained on the entire space dataset. 
\subsubsection{Reinforcement Learning} 
Reinforcement learning (RL) has emerged as a powerful tool in cross-domain recommendation, offering a dynamic approach to optimizing recommendation strategies by learning from user interactions and adapting to user preferences across multiple domains. 
DACIR \citep{wu2022dynamics} represents an early attempt to leverage reinforcement learning in cross-domain recommendation by introducing a reward function to calibrate dynamic interaction patterns across two domains, effectively narrowing the cross-domain gap.

Unlike traditional methods, which often rely on static or pre-defined weights and feature selections, RL-based frameworks can adaptively adjust their strategies to maximize long-term user engagement and satisfaction.
AutoTransfer \citep{gao2023autotransfer} introduces a reinforcement learning algorithm to train the policy network so that it can select the best action based on the current state. At the same time, in order to further improve the generalization ability of the model, AutoTransfer also introduces a data reuse mechanism so that the selected instances can be used in subsequent training of different models.
However, user preferences and domain characteristics may vary greatly, requiring dynamic and context-aware adjustments.
FREMIT \citep{sun2023remit} framework uses reinforcement learning to dynamically adjust the weights assigned to transformed interests for different training instances. This allows the framework to adaptively select the most relevant interests for each user and item combination, thus improving the performance of the target model. The reinforcement learning algorithm is based on the task-oriented optimization approach, which directly utilizes the performance of the ultimate recommendation task as the optimization goal.
\subsection{Decentralized Training} \label{dct}
Cross-domain recommendation systems need to be able to generalize to different domains and user groups. Decentralized architectures enhance the robustness and scalability of the system. In cross-domain recommendation systems, decentralization allows each node to learn independently, reduces the risk of single point failure, and can better adapt to the specific needs of different domains. 
Specifically, federated learning and privacy protection technologies are widely used in cross-domain recommendation.

\subsubsection{Federated Learning}
Federated learning enables multiple domains to collaboratively train a recommendation model without directly sharing their data. 
PFCR \citep{guo2024prompt} utilizes a federated learning schema by exclusively using users' interactions with local clients and devising an encryption method for gradient encryption. The authors also model items in a universal feature space by their description texts and initially learn federated content representations, harnessing the generality of natural language to establish bridges between domains.
By transferring knowledge from aligned data to unaligned data,
FedUD \citep{ouyang2024fedud} presents an innovative approach to address the limitations of traditional vertical federated learning by incorporating unaligned data into the training process while preserving privacy.
\subsubsection{Privacy Protection }
Privacy issues in cross-domain recommendation have gradually received attention from researchers, including how to protect users' personal information and behavior data from being leaked, and how to achieve effective cross-domain recommendation while protecting privacy. 
FPPDM \citep{liu2023federated} addresses challenges associated with privacy protection and data sparsity in multi-domain scenarios. The local domain modeling component focuses on exploiting user/item preference distributions within each domain using available rating information. On the other hand, the global server aggregation component aims to combine user characteristics across domains, enabling effective extraction of semantic neighbor information among users.
P2DTR \citep{linenhancing} introduces a novel inter-client knowledge extraction mechanism leveraging the private set intersection algorithm and prototype-based federated learning to enable collaborative modeling, while DeCoCDR \citep{li2024decocdr} tackles similar privacy issues by dividing the recommendation process into two stages: cloud-based recall and local re-ranking.

\subsection{Other Optimization Methods}

Many researchers have made numerous attempts to optimize models in cross-domain recommendation, among which graph-based approaches have shown significant potential by leveraging graph structures to model complex relationships between users and items across domains. These methods utilize innovative graph representation and embedding techniques to enhance recommendation accuracy by capturing high-order and cross-domain interactions.
GReS \citep{jing2022gres}, designed for catering supply platforms, enhances recommendation performance by using a tree-shaped graph structure to represent hierarchical relationships among nodes. It embeds users and items with Node2vec and leverages the Tree2vec method—combining GCN and BERT models—to embed the tree-shaped graph effectively.
Building on the concept of capturing intra-domain relationships, II-HGCN \citep{han2023intra} constructs user and item hypergraphs to explore high-order user relationships within each domain. By using hypergraph convolution in the intra-domain layer, it updates embeddings, and further builds an inter-domain hypergraph based on common users.
Extending these ideas to a multi-domain setting, H3Trans \citep{xu2023correlative} constructs a unified multi-domain graph with two hyper-edge-based modules: dynamic item transfer (Hyper-I) and adaptive user aggregation (Hyper-U). Hyper-I identifies similar items in the target domain for each source domain item and connects them via hyper-edges, while Hyper-U utilizes high-order connections across users’ scattered preferences in different domains to improve user representations comprehensively.

Meanwhile, some researchers believe auxiliary tasks play a crucial role in enhancing cross-domain recommendation models by providing additional learning signals to improve representation quality and optimize information transfer. CCTL \citep{zhang2023collaborative} utilizes a symmetric companion network to evaluate the information gain from source domains. It adjusts information transfer weights through an information flow network while employing a representation enhancement network as an auxiliary task to retain domain-specific features. Expanding the scope of auxiliary tasks, CDIMF \citep{samra2024cross} employs ADMM, an algorithm designed for solving constrained convex optimization problems, to introduce auxiliary and dual variables. By treating the alignment of user factors from different domains as a constraint-based auxiliary task, CDIMF ensures that the user representations remain consistent across domains.


Finally, some researchers also try to improve the efficiency of the recommendation model. AdaSparse \citep{yang2022adasparse} addresses the challenge of improving generalization across domains under limited training data and reducing computational complexity. By learning adaptively sparse structures for each domain, AdaSparse achieves better generalization across domains with lower computational cost compared to existing multi-domain models.


\begin{table*}[t]
  \centering
  \caption{Summary of the CDR Datasets}
  \resizebox{1\textwidth}{!}{
    \begin{tabular}{cccc}
    \toprule
    Datasets & Domains  & Scale & Link \\
    \midrule
    \midrule
    Amazon & 
     \begin{tabular}{cc}
  Movie, Book, Music, & \\ Sport, Phone, Elec, Cloth, CD 
    \end{tabular}
    & 1w & {http://jmcauley.ucsd.edu/data/amazon/index\_2014.html} \\
        \midrule
    Douban & Movie, Book, Music   & 1M & {https://recbole.s3-accelerate.amazonaws.com/CrossDomain/Douban.zip} \\
        \midrule
    Tenrec & 
     \begin{tabular}{cc}
  QK-article, QK-video, & \\ QB article, QB video  
    \end{tabular}
      & 100M + & {https://static.qblv.qq.com/qblv/h5/algo-frontend/tenrec\_dataset.html} \\
        \midrule
    HVIDEO & V-domain, E-domain  & 1M & {https://bitbucket.org/Catherine\_Ma/pinet\_sigir2019/src/master/HVIDEO}/  \\
        \midrule
    Taobao & 
    \begin{tabular}{cc}
 Thematic topics, & \\ such as what to take when traveling & \\ how to dress up yourself for a party &\\ things to prepare when a baby is coming
    \end{tabular}
     & 1.4M & 
    {https://tianchi.aliyun.com/dataset/9716} \\
        \midrule
    AliCCP &  Positions of Alimama’s traffic logs     & 20M & 
    {https://tianchi.aliyun.com/dataset/408} \\
        \midrule
    AliAd & Ads categories  & 26M & 
    {https://tianchi.aliyun.com/dataset/56} \\
    \bottomrule
    \end{tabular}}%
  \label{tab:dataset}%
\end{table*}%

\section{Application \& Resource}
To address the challenge of heterogeneity in real industrial settings, \cite{huan2023samd}  propose an industrial framework to resolve the heterogeneity in multi-scenario recommendations, which has been applied to Alipay's advertising system.
To tackle the data sparsity problem in Supply Chain Platforms (SCPs), \cite{jing2022gres} introduce GReS, a graphical Cross-Domain Recommendation (CDR) model. This model constructs a tree-shaped graph to represent the hierarchy of different nodes of dishes and ingredients, and then combines Graph Convolutional Networks (GCN) and BERT models to embed the graph for recommendations. And \cite{wang2024pre} proposes a novel pre-train then fine-tune framework specifically for cross-market recommendation, which introduces a novel pre-training function to address item popularity market shift and mitigate its impact on recommendations.
Considering that existing CDR models fail to enhance recommendation performance across different domains in real-world applications where items in the source domain are entirely different from those in the target domain in terms of attributes. \cite{zheng2023dual} propose the concept of inter-domain interest alignment, supported by a convincing analysis of different domains in WeChat. Additionally, we have noticed that some other work \citep{xie2022contrastive} has also been applied to cross-domain recommendation scenarios within WeChat.

For cross-domain recommendation research, datasets play a crucial role, particularly in advancing studies in these specific applications. To facilitate access to these essential resources, we summarize several popular cross-domain recommendation datasets by application in Table \ref{tab:dataset}, providing researchers with an accessible guide for locating these datasets. Researchers interested in further details may refer to the relevant literature and resources provided.

\begin{itemize}

\item
Amazon is a comprehensive e-commerce dataset containing customer reviews and ratings from over 2 million users. The dataset spans across multiple product categories, such as electronics, books, fashion, and home appliances. Each interaction includes detailed information such as review text, rating (on a 5-star scale), product ID, and user ID, providing a rich resource for analyzing customer behavior, sentiment, and product performance.
\item 
Douban is a comprehensive dataset derived from Douban, a Chinese social networking service focusing on books, movies, and music. The dataset contains rich information, such as ratings (on a 5-star scale), reviews, tags, user profiles, and item details.

\item
Tenrec \citep{yuan2022tenrec} is sourced from Tencent’s recommendation platforms, with each item being either a news article or a video. It encompasses various forms of positive user feedback, such as clicks, likes, shares, follows, reads, and favorites, while also including true negative feedback for comprehensive analysis.
\item
HVIDEO is a TV dataset containing viewing logs from 260,000 users, recorded from October 1, 2016, to June 30, 2017. The logs are collected across two platforms, referred to as the V-domain and the E-domain, from a prominent smart TV service provider.
\item 
Taobao \citep{taobao_du2019sequential} is derived from the click logs of the Cloud Theme in Taobao app, which tailors item collections to specific scenarios based on user interests. It contains over 1.4 million clicks from 355 scenarios during a 6-day promotion period, along with one month of users’ purchase history before the promotion started.
\item 
AliCCP \cite{aliccp_ma2018entire} consists of traffic logs collected from Taobao's recommender system, with a 1\% random sample made publicly available. And the data are often divided into different domains according to positions.

\item
AliAd is an advertising dataset provided by Alibaba, containing randomly sampled ad display and click logs from 1.14 million Taobao users over an 8-day period, with a total of 26 million records. User interactions with advertised products include browsing, adding to cart, favoring, and purchasing.
\end{itemize}
\section{Challenges and Future Directions}
Despite the effectiveness of the existing works, cross domain recommender system still face several challenges.
\begin{itemize}
    \item \textbf{Negative Transfer}. Numerous existing techniques indiscriminately transfer all information from source domains to the target domain. This process can introduce detrimental noise and irrelevant features, leading to less than optimal performance. This occurrence, known as negative transfer, has been documented in various previous works \citep{song2024mitigating,xu2021expanding,zhao2022multi}, and still needs to be explored.
    \item \textbf{Data Imbalance}. The distribution of data across different domains can be uneven, which can lead to biased learning and poor recommendation performance. This imbalance in data distribution is a persistent issue that needs further investigation.
    \item \textbf{Recommendation Interpretability}.While many existing methods can provide accurate recommendations, they often lack interpretability. It is important for users to understand why certain items are recommended under the cross domain scenarios, which is a challenge that needs to be addressed.
    \item \textbf{Privacy}. Cross domain recommendation often involves sharing user data across different domains or platforms, which raises privacy concerns. Ensuring privacy while maintaining recommendation quality is a significant challenge in this area.
\end{itemize}
In addition, we also provide some future directions:
\begin{itemize}
    \item \textbf{Utilization of Large Models}. Large language model (LLM) is known for its remarkable capability of language understanding and reasoning
in recent years. Although there has been some work attempting to improve cross-domain recommendation performance using LLMs, many issues remain to be explored. For example, how to enable LLMs to understand domain-specific knowledge, how to infuse cross-domain interaction information into LLMs, and how can we address the efficiency issues of LLMs in cross-domain recommendation scenarios, and so on.

    \item \textbf{Cross-Domain Recommendation with Lifelong Learning}.
  Lifelong learning seeks to continuously learn from an infinite stream of data, progressively building upon previously acquired knowledge to enhance future learning. A primary focus of this paradigm is addressing the challenge of catastrophic forgetting. However, the application of lifelong learning in cross-domain settings remains largely unexplored in both academic research and industrial practice.

    \item \textbf{Multimodal Data Fusion and Understanding}. Multimodal data fusion has shown great potential in improving recommendation systems by integrating diverse types of data, such as text, images, audio, and video. While some studies have explored utilizing multimodal data for cross-domain recommendation, numerous challenges remain unresolved. For instance, how to effectively extract and align features across different modalities, how to handle the heterogeneity and noise inherent in multimodal data, and how to design scalable models that can process large-scale multimodal inputs efficiently in real-world scenarios.

    \item \textbf{Cross-Cultural and Multilingual Recommendation}. Cross-cultural and multilingual recommendation is gaining attention as global platforms aim to cater to diverse user bases. Although initial efforts have been made to address the unique challenges in this area, many open questions remain. For example, how to effectively model cultural nuances and regional preferences, how to bridge the semantic gaps caused by linguistic differences, and how to design scalable algorithms that can adapt to dynamic and diverse cultural contexts. Moreover, addressing the sparsity of user interaction data in less-dominant languages and incorporating contextual and cultural knowledge into recommendation models are critical challenges yet to be fully explored.
\end{itemize}
    
\vspace{0.3in}
\section{Conclusion}
Cross-Domain Recommendation (CDR) is a widely used approach
for leveraging information from domains to achieve better recommendation performance. In this paper, we summarized past work from a technical perspective based on cross-domain rec-
ommendation procedures.
Besides, this article also includes
the latest advancements such as LLM-based
approaches and cross-domain representation enhancement methods. Furthermore, we listed several popular CDR datasets and real-world applications, and summarized the current challenges faced by CDR as well as future directions.
We hope this paper could
inspire more works to be proposed 
for cross domain recommendation.




\bibliographystyle{cas-model2-names}

\bibliography{main}


\end{document}